\documentstyle[12pt]{article}
\def\beq{\begin{equation}}
\def\eeq{\end{equation}}
\def\bea{\begin{eqnarray}}
\def\eea{\end{eqnarray}}
\def\bq{\begin{quote}}
\def\eq{\end{quote}}
 
\begin{document}
\pagestyle{empty}
\begin{flushright}
{ROME prep.1249/99 \\
 hep-th/9904200}
\end{flushright}
\vspace*{5mm}
\begin{center}
{\bf THE ABELIAN PROJECTION VERSUS THE HITCHIN FIBRATION OF
$K(D)$ PAIRS IN FOUR-DIMENSIONAL $QCD$}
\\  
\vspace*{1cm} 
{\bf Marco Bochicchio} \\
\vspace*{0.5cm}
INFN Sezione di Roma \\
Dipartimento di Fisica, Universita' di Roma `La Sapienza' \\
Piazzale Aldo Moro 2 , 00185 Roma  \\ 
\vspace*{1cm} 
\vspace*{0.5cm}
\vspace*{2cm}  
{\bf ABSTRACT  } \\
\end{center}
\vspace*{5mm}
\noindent
We point out that the concept of Abelian projection gives us a physical
interpretation of the role that the Hitchin fibration of parabolic $K(D)$
pairs plays in the large-$N$ limit of four-dimensional $QCD$. \\ 
This physical interpretation furnishes also a simple criterium for
the confinement of electric fluxes in the large-$N$ limit of $QCD$. \\
There is also an alternative, compatible interpretation, based on the
$QCD$ string.
\vspace*{1cm}
\begin{flushleft}
April 1999
\end{flushleft}
\phantom{ }
\vfill
\eject

\setcounter{page}{1}
\pagestyle{plain}

\section{Introduction}

Some years ago 't Hooft introduced the concept of Abelian projection
\cite{H1} into non-Abelian gauge theories, in order to explain the confinement
of quarks in four-dimensional $QCD$ as a dual Meissner effect in a dual
superconductor \cite{H02, M}. \\
The Abelian projection allows us, by a careful choice of the gauge, to 
describe the physical variables of a non-Abelian $SU(N)$ gauge theory, without 
scalar matter fields, as a set of electric charges and magnetic monopoles 
interacting via a residual $U(1)^{N-1}$ Abelian gauge coupling. \\
The occurrence of magnetic monopoles into a  non-Abelian gauge theory
without matter fields is perhaps the most crucial feature of the Abelian 
projection, that furnishes a precise understanding of the structure of the 
phases of non-Abelian gauge theories, according to the following 
alternatives \cite{H3}. \\
If there is a mass gap, either the electric charge condenses in the vacuum
(Higgs phase) or the magnetic charge does (confinement phase).
If there is no mass gap, the electric and magnetic fluxes coexist
(Coulomb phase). \\
Recently, in an apparently unrelated development \cite{MB}, some mathematical 
control was gained over the large-$N$ limit of four-dimensional $QCD$, mapping,
by means of a chain of changes of variables, the function space of the $QCD$ 
functional integral into an elliptic fibration of Hitchin bundles. \\
Hitchin bundles \cite{Hi} are themselves a fibration of $U(1)$ 
bundles over spectral branched covers of a Riemann surface, that, in the case 
of \cite{MB}, is a torus. \\
In this paper, we point out that the map in \cite{MB} is a version,
in a perhaps global algebraic-geometric setting, of the concept of Abelian 
projection \cite{H1}. \\ 
In fact, the branching points of the spectral cover are identified with the 
magnetic monopoles of the Abelian projection, the parabolic points of the cover
with (topological) electric charges and the $U(1)$ gauge group on the cover 
with a global version (on the cover) of the $U(1)^{N-1}$ gauge group of the 
Abelian projection. \\
The identifications that we have just outlined provide a physical 
interpretation of the mathematical construction in \cite{MB}.
Indeed it is precisely this physical interpretation
that explains naturally why the functional integral, once it is expressed as a 
functional measure supported over the collective field of the Hitchin fibration,
is dominated by a saddle-point condition in the large-$N$ limit. \\
On the other side, we may think that the mathematical proof, that the 
variables of the Abelian projection really capture the 
physics of four-dimensional $QCD$ in the large-$N$ limit, relies on the fact 
that those variables may be employed to dominate the functional integral in 
the large-$N$ limit. \\
The only qualitative feature, in the treatment in \cite{MB}, that was not 
already present in the concept of the Abelian projection, is the occurrence
of Riemann surfaces and it is due to the global algebraic-geometric nature
of the methods in \cite{MB}. This, however, makes contact, at least 
qualitatively, with another long-standing conjecture about the $QCD$ 
confinement, the occurrence of string world sheets \cite{GT} and the string 
program \cite{Po}. \\
Our last concluding remark is that the electric/magnetic alternative
\cite{H3} and the physical interpretation based on the Abelian projection,
applied in the mathematical framework of \cite{MB}, give us a simple
qualitative criterium to characterize the confinement phase of $QCD$ in the 
large-$N$ limit: confinement is equivalent to magnetic condensation, in absence 
of electric (parabolic) singularities of the spectral covers. \\
An alternative, compatible interpretation, based on the idea that $QCD$
is equivalent, in the large-$N$ limit, to a theory of strings \cite{GT,Po}
is outlined in the following section.
The rest of the paper is devoted to a technical explanation of the 
correspondence between the Abelian projection and the Hitchin fibration
in four-dimensional $QCD$.

\section{The Hitchin fibration as the Abelian projection in the gauge in which
the Higgs current is a triangular matrix}

The Abelian projection, according to \cite{H1}, is really the choice of a 
gauge-fixing in such a way that, after the gauge-fixing, the theory is no 
longer locally invariant under $SU(N)$ but only under its Cartan subgroup 
$U(1)^{N-1}$. The important point about this projection is that it is defined 
strictly locally, that is, the gauge rotation $\Omega$ performed at each point 
in space-time to implement the gauge-fixing condition, does not depend on the 
values of the physical fields in other points of space-time.
This then guarantees that all observables in the new
gauge frame are still locally observable.  There are no propagating ghosts.
But $\Omega$ is not completely defined.  There is a subgroup,
$U(1)^{N-1}$, of gauge rotations that may still be performed.  And
this is why the theory, after the Abelian projection, looks like a local
$U(1)^{N-1}$ gauge theory. \\
If one now tries to gauge-fix this remaining gauge freedom, one
discovers that it cannot be done locally, without encountering
apparent difficulties. But local gauge-fixing is not needed, since
the residual gauge symmetry is the one of a familiar Abelian theory. \\
There may be, however, isolated points, where the local gauge-fixing condition
has coinciding eigenvalues, where the gauge symmetry is not $U(1)^{N-1}$ but a 
larger group.  Here singularities appear, the magnetic monopoles.
So we see that, topologically, the full theory can only be
topologically equivalent to the $U(1)^{N-1}$ gauge theory if the
latter is augmented with monopole singularities where the $U(1)$
conservation laws for the vortices are broken down into the (less restrictive)
conservation laws of the $SU(N)$ vortices. \\
When we try to gauge-fix completely, we hit upon the Dirac strings, whose
end points are the magnetic monopoles. \\
In addition to the magnetic monopoles, in the $QCD$ case, the gauge-fixed 
theory contains also gluon and quark fields, that are charged with respect to 
the residual $U(1)^{N-1}$. \\
Therefore we have a set of electric charges and magnetic monopoles 
interacting via a residual $U(1)^{N-1}$ Abelian gauge coupling. \\
We now compare this description with the one that arises in \cite{MB}, for
the pure gauge theory without quark matter fields. \\
The functional integral for $QCD$ in \cite{MB} is defined in terms of the
variables $(A_z, A_{\bar z}, \Psi_z, \Psi_{\bar z})$, obtained by means of a 
partial duality transformation from $(A_z, A_{\bar z}, A_u, A_{\bar u})$, 
where $(z, \bar z, u, \bar u)$ are the complex coordinates on the product of 
two two-dimensional tori, over which the theory is defined. \\
$(A_z, A_{\bar z}, \Psi_z, \Psi_{\bar z})$ define the coordinates of an 
elliptic fibration of $T^* {\cal A}$, the cotangent bundle of unitary 
connections on the $(z, \bar z)$ torus, whose base is the $(u, \bar u)$ torus.
\\
$\Psi_z$ transforms as a field strength by gauge transformations
and it is a non-hermitian matrix. \\
Following Hitchin \cite{Hi}, the gauge is chosen in which $\Psi_z$ is a 
triangular matrix, for example lower triangular, that leaves a $U(1)^{N-1}$ 
residual gauge freedom as in the Abelian projection. \\
The points in space-time where $\Psi_z$ has a pair of coinciding eigenvalues,
correspond to monopoles.
In addition there are the charged components of 
$(A_z, A_{\bar z}, \Psi_z, \Psi_{\bar z})$. 
We have thus a set of charges and monopoles with a residual $U(1)^{N-1}$,
according to the Abelian projection. \\
In \cite{MB}, however, it is found a dense set in the functional 
integral over (the elliptic fibration of) $T^* {\cal A}$, with the property that
the quotient by the action of the gauge group exists as a Hausdorff (separable)
manifold. \\
This dense set is defined in \cite{MB} as the set of 
pairs $(A, \Psi)$ that are solutions of the following differential equations
(elliptically fibered over the $(u, \bar u)$ torus):
\bea
F_A-i \Psi \wedge \Psi &=& \frac{1}{|D|}\sum_p \mu^{0}_{p} \delta_p
i dz \wedge d\bar{z} \nonumber \\
\bar{\partial}_A \psi &=& \frac{1}{|D|}\sum_p \mu_{p} \delta_p
dz \wedge d\bar{z} \nonumber \\
\partial_A \bar{\psi} &=& \frac{1}{|D|}\sum_p \bar{\mu}_{p} \delta_p
d\bar{z} \wedge dz \;
\eea
where $\delta_p$ is the two-dimensional delta-function localized at $z_p$ 
and $(\mu^{0}_{p},\mu_{p},\bar{\mu}_{p})$ are the set of levels for the 
moment maps. The moment maps are the three Hamiltonian densities generating 
gauge transformations on $T^* {\cal A}$ that appear in the left hand sides of 
Eq.(1) \cite{Hi1}. \\
$\mu^{0}_{p}$ are hermitian traceless matrices, and $\mu_{p}$ are matrices in 
the complexification of the Lie algebra of $SU(N)$, that 
determine the residues of the poles the Higgs current $\Psi$.
$\psi$ and $\bar{\psi}$ are the $z$ and $\bar z$ components of the one-form 
$\Psi$. \\
Eq.(1) defines a dense stratification of the functional integral over
$T^* {\cal A}$ because the set of levels is dense everywhere in function 
space, in the sense of the distributions, as the divisor $D$ gets larger and 
larger. \\
Eq.(1) defines the data of parabolic $K(D)$ pairs \cite{K} on a torus valued 
in the Lie algebra of the complexification of $SU(N)$: a holomorphic 
connection $\bar{\partial}_A$ of 
a holomorphic bundle, $E$, with a parabolic structure and a parabolic morphism 
$\psi$ of the parabolic bundle. The parabolic structure at a point $p$ 
\cite{MS,K} 
consists in the choice of a set of ordered weights, that are positive real 
numbers modulo 1, and a flag structure, that 
is a collection of nested subspaces $ \cal{F}_{1} 
\subset \cal{F}_{2} \subset...\cal{F}_{i}$ labelled by the weights $\alpha_1 \geq \alpha_2 
\geq ...\alpha_k$, with the
associated multiplicities defined as: $m_{i+1}=dim \cal{F}_{i+1}-dim 
\cal{F}_{i}$.
A parabolic morphism, $\phi$, is a holomorphic map between parabolic 
bundles,
$E^1,E^2$, that preserves the parabolic flag structure at each parabolic 
point $p$ in the sense that $\alpha^{1}_{i} > \alpha^{2}_{j}$ implies 
$\phi( \cal{F}^{1}_{i}) \subset \phi(\cal{F}^{2}_{j+1})$.
We should now explain how a parabolic structure arises from Eq.(1) and how 
it follows that $\psi$ is a parabolic morphism with respect to the given
parabolic structure.
Though we are going to choose the gauge in which $\psi$ is a lower triangular
matrix in 
most of this paper, we start at an intermediate stage with a 
gauge in which $\mu^{0}_{p}$ is diagonal. The eigenvalues of $\mu^{0}_{p}$
modulo $2 \pi$ and divided by $2 \pi$ define the parabolic weights.
Their multiplicities will turn out to be the multiplicities of the yet to 
be defined flag structure.\\
Fixed $\mu^{0}_{p}$ and $\mu_p$ in Eq.(1), let $(e_k)$ be an orthonormal
basis of the eigenvectors of $\mu^{0}_{p}$ in decreasing order. This basis 
is not necessarily unique if the eigenvalues have non-trivial 
multiplicities. However the corresponding flag structure will not be affected 
by this lack of uniqueness.
Let $g$ be
the gauge transformation that puts $\mu$ and $\psi$ into lower 
triangular form. Let $(g e_k)$ be
the transformed basis and let $\cal F$ be the flag obtained by taking the unions 
of subspaces generated by the vectors in the transformed basis 
that are the images of eigenvectors of the ordered 
eigenvalues with the given multiplicity in such a way that the multiplicities
of the resulting flag are the same as the multiplicities of 
the eigenvalues. In addition, by construction, $\psi$ is a parabolic morphism
with respect to the flag since it is holomorphic and lower triangular in the 
basis $(ge_k)$. \\
We have thus the data of a parabolic $K(D)$ pair from Eq.(1). \\
There is also a representation theoretic interpretation of Eq.(1). \\
The three equations for the moment maps are equivalent to a vanishing 
curvature condition for the non-hermitian connection one-form $B=A+i \Psi$ plus
a harmonic condition for $\psi$ away from the parabolic 
divisor \cite{S}. \\
Therefore the set of solution of Eq.(1) can be figured out essentially as a 
collection of monodromies around the points of the divisor with values in the 
complexified gauge group, that form a representation of the fundamental group 
of the torus with the points of the parabolic divisor deleted. \\
't Hooft description of the Abelian projection previously outlined,
applies to $T^* {\cal A}$ and to its dense subset defined by Eq.(1)
a fortiori. In addition, we have just shown that there is an embedding of the 
solutions of Eq.(1) into the parabolic $K(D)$
pairs. \\
However, on the parabolic $K(D)$ pairs, 't Hooft concept of Abelian projection
can be carried to its extreme consequences. \\
Indeed, in the global algebraic-geometric framework of the Hitchin fibration
\cite{Hi,K} of parabolic $K(D)$ pairs, it is preferable to concentrate 
ourselves on the first eigenvalue and the first eigenstate of the lower 
triangular matrix $\Psi_z$, since all the information of the original 
parabolic bundle, up to gauge equivalence, can be reconstructed from these 
only data \cite{Hi}. \\
The first eigenvalue defines a spectral covering, that is a branched cover
of the two-torus. The eigenspace defines a section of a line bundle, that 
determines a $U(1)$ connection on the cover of the torus, instead of the 
$U(1)^{N-1}$ bundle on the torus of the Abelian projection. \\
The $U(1)$ connection on the cover, $a$, and the eigenvalue, $ \lambda$, of the 
Higgs current can be considered as coordinates of the cotangent bundle of 
unitary $U(1)$ connections on the cover, or as parabolic $K(D)$ 
pairs $(a, \lambda)$ on the cover, valued in the complexification of the Lie 
algebra of $U(1)$. \\
The system is now completely abelianized. Correspondingly, not only
the magnetic charges, but also the electric ones can occur only as gauge 
invariant topological configurations. \\
The points in space-time where $\Psi_z$ has a pair of coinciding eigenvalues,
that in the Abelian projection correspond to monopoles, are here, 
according to Hitchin, simple branching points of the spectral covers, defined 
by means of the characteristic equation: 
\bea
Det(\lambda 1-\Psi_{z})=0 , 
\eea
in which the coordinates $(u,\bar{u})$ are kept fixed. \\
All the other branching points can be obtained by collision of these simple
branching points, in the same way monopoles can in the Abelian projection.
The branching points are the end points of string cuts on the Riemann 
surfaces, the Dirac strings of the Abelian projection. \\
These Riemann surfaces, the only additional global ingredient with respect
to the Abelian projection, are interpreted as the world sheets of  
strings made by electric flux lines. \\
A closed string of electric flux is represented by a Wilson loop of the 
$U(1)$ connection $a$ on the cover, along a non-trivial generator of the 
fundamental group of the surface. \\
In addition, the Riemann surfaces, defined by the spectral equation,
may posses parabolic points, associated to poles of the eigenvalues
of the Higgs current $\Psi_z$, whose origin is in the parabolic 
singularities of the original $su_{c}(N)$-valued $K(D)$ pair, which may be 
reflected into a parabolic structure for the $u_{c}(1)$-valued $K(D)$ pair on the 
cover. \\ 
These poles, together with the ones of the $U(1)$ connection, are interpreted 
as electric charges. Indeed it is not difficult to see that they are electric
sources, that appear where a boundary-electric loop shrinks to a point. \\
Therefore, the electric charges occur here as topological objects associated
to the parabolic degree \cite{MS} of the $u_{c}(1)$-valued $K(D)$ pair.
On the other side, magnetic topological quantum numbers are associated,
as usual, to the ordinary degree of the $U(1)$ bundle. \\
We should mention however that a subtlety arises in our interpretation
of the Hitchin fibration in terms of the Abelian projection.
As we mentioned in the first part of this section, in the Abelian projection
the gauge-fixing condition leaves a residual non-Abelian gauge symmetry
where a magnetic monopole occurs. This is essentially due to the fact that 't 
Hooft chooses to diagonalize a hermitian functional of the fields. On the 
contrary, in the case of the dense set 
defined by Eq.(1), since $\psi$ is a non-hermitian matrix, it can only be put
in triangular form. This gauge-fixing does not leave in general
a residual compact non-Abelian gauge symmetry even when the eigenvalues 
coincide.
However this difficulty can be resolved in the following way, anticipating
somehow some of the conclusions of this paper and the result of \cite{MB2}.
Let us require for the moment that the levels of the non-hermitian moment maps
be nilpotent. Since these are only $N$ conditions at each parabolic point they 
do not modify essentially the entropy of the functional integration in the 
large-$N$ limit. The true physical meaning of this choice has to do with 
confinement and it is explained in \cite{MB2}.
If the residues of the Higgs field are nilpotent, Eq.(1) can be interpreted 
as the vanishing condition for the moment maps of the action of the compact 
$SU(N)$ gauge group on the pair $(A, \Psi)$ and on the cotangent space of 
coadjoint orbits \cite{Ale}:
\bea
&&F_A-i \Psi \wedge \Psi - \frac{1}{|D|}\sum_p \mu^{0}_{p} \delta_p
i dz \wedge d\bar{z}=0 \nonumber \\
&&\bar{\partial}_A \psi- \frac{1}{|D|}\sum_p n_{p} \delta_p
dz \wedge d\bar{z}=0 \nonumber \\
&&\partial_A \bar{\psi}- \frac{1}{|D|}\sum_p \bar{n}_{p} \delta_p
d\bar{z} \wedge dz=0 \;
\eea
In addition the quotient under the action of the compact gauge group is 
hyper-Kahler \cite{K}. By a general result of Hitchin, Karlhede, Lindstr\"{o}m and
Roc\v{e}k \cite{H2}, the hyper-Kahler quotient under the 
action of the compact gauge group in Eq.(3) is the same as the quotient 
defined by the non-hermitian moment maps:
\bea
&&\bar{\partial}_A \psi- \frac{1}{|D|}\sum_p n_{p} \delta_p
dz \wedge d\bar{z}=0 \nonumber \\
&&\partial_A \bar{\psi}- \frac{1}{|D|}\sum_p \bar{n}_{p} \delta_p
d\bar{z} \wedge dz=0 \;
\eea
under the action of the complexification of the gauge group.
We can therefore impose a gauge condition 
compatible with the compact action in Eq.(3) or a gauge condition compatible 
with the action of the complexified group in Eq.(4) getting the same moduli 
space.
In the second case we choose the gauge in which
$\psi$ is diagonal. This condition becomes singular
where two or more eigenvalues coincide. In fact it cannot be extended 
continuously to the points where the eigenvalues coincide. There it can only 
be required that $\Psi_z$ be a triangular matrix. However this condition 
leaves now a residual non-Abelian gauge symmetry in the complexification of the gauge 
group: the freedom of making triangular gauge transformations, thus 
confirming our analogy with 't Hooft definition of magnetic monopoles. \\
To summarize, the ingredients of the Hitchin fibration of the $su_{c}(N)$-
valued $K(D)$ pairs, are the branching points, that are interpreted as 
magnetic monopoles, and the $U(1)$ monodromies around closed loops, that are interpreted
as electric lines. In addition, the ordinary degree of the $U(1)$ 
bundle is interpreted as a (topological) magnetic charge, while the parabolic 
degree \cite{MS} of the $U(1)$ bundle is interpreted as a (topological) electric charge.
\\
The difference here, with the letter but not with the spirit of the Abelian 
projection, is that the system has been completely abelianized, so 
that both the magnetic and the electric charges are topological.
We are thus given a set of charges and monopoles with a $U(1)$ gauge group on 
the covering, in analogy with the Abelian projection. \\
We call this description a complete Abelian projection. \\
The string interpretation is as follows. 
The spectral covers are the world sheets of strings, made by the electric flux 
lines. The confinement condition is equivalent to requiring
that only closed string world sheet occur, since confinement requires
that the flux lines can never break in absence of quarks. \\
If the spectral covers posses parabolic points, the same as 
electric charges in the complete Abelian projection, they are, 
topologically, Riemann surfaces with boundaries at infinity. \\
For example a sphere with two parabolic points is a topological cylinder. \\
But a cylinder can occur in vacuum string world sheets (we are describing 
the contributions to the partition function, the vacuum to vacuum amplitude 
indeed) only if open strings propagate. \\
In fact, a closed string that propagates through the torus
breaks into an open one at the parabolic points, since the parabolic points do
not belong to the world sheet. \\
On the contrary, when a closed string meets a 
branching point, for example in a once-branched double cover of a torus, the 
closed string is  pinched into another closed string with the form of a double
loop intersecting at the (simple) branching point. \\
Notice also that the branching points do belong to the world sheet. \\
Thus, the string picture is consistent with the interpretation
of branching points as magnetic charges, where the string electric line can
self-intersect but not break, and of parabolic points as electric 
charges, where closed string break into open strings with the parabolic points 
as boundaries.

\section{Conclusions}

Our conclusion is that the concept of Abelian projection in \cite{H1} furnishes
a physical interpretation of the structures that appear in the
Hitchin fibration of $K(D)$ pairs, as it is embedded in the $QCD$
functional integral in \cite{MB}. \\
In addition, there is a complementary consistent string interpretation. \\
The most relevant consequence of these interpretations is a criterium
for electric confinement in the framework of \cite{MB}, that is the usual
criterium of magnetic condensation of \cite{H1}.\\
Therefore, if $QCD$ confines the electric charge, the functional measure must 
be localized, in the large-$N$ limit, on those parabolic $K(D)$ pairs, whose 
image through the Hitchin map, contains monopoles but no charges, that is, in 
geometric language, on those spectral covers that are arbitrarily branched, 
but that do not posses a parabolic divisor. \\
In turn, this is equivalent to the condition that only spectral covers 
spanned by closed strings occur as configurations in the vacuum to vacuum
amplitude. \\
It is amusing to notice that this condition is satisfied by the string
of two-dimensional $QCD$ in the large-$N$ limit \cite{GT}.

\section{Acknowledgements}

We would like to thank Gerard 't Hooft for several clarifying remarks
on the Abelian projection.

\end{document}